\documentclass[a4paper,10pt]{article}

\usepackage{contribution}
\usepackage{subfig}
\usepackage{caption}
\newcommand{\weblink}[2][]{%
    \ifthenelse{\equal{#1}{}}%
    {\textnormal{\url{#2}}}%
    {\textnormal{\href{#2}{#1}}}%
}

\def\beq{\begin{equation}}
\def\eeq#1{\label{#1}\end{equation}}
\def\eeqn{\end{equation}}

\def\beqa{\begin{eqnarray}}
\def\eeqa#1{\label{#1}\end{eqnarray}}
\def\eeqan{\end{eqnarray}}

\let\bar=\overbar
\providecommand{\keywords}[1]{\textit{Keywords---} #1}

\def\Dslash{\not{\hbox{\kern-4pt $D$}}}
\def\dslash{\not{\hbox{\kern-2pt $\del$}}}

\def\msb{{\bar{\ssstyle M \kern -1pt S}}}

\newcommand{\contribution}[7][]{%
  \clearpage
  \thispagestyle{plain}

  \ifthenelse{\equal{#1}{}}
  {\hypersetup{pdftitle={#2}}}
  {\hypersetup{pdftitle={#1}}}
  \hypersetup{pdfauthor={{#3} {#4}}}
  {\centering\normalfont\LARGE\bfseries\sffamily #2 \par\nobreak}
  \lhead{}
  \chead{%
    \textit{\footnotesize XXIInd International Workshop ``High-Energy Physics and Quantum Field Theory'', 
June 24 -- July 1, 2015, Samara, Russia}%
  }
  \rhead{}
  \bigskip
  \begin{center}
    {#3} {#4}\ifthenelse{\equal{#6}{}}{}{\footnote{\weblink[#6]{mailto:#6}}}
    \ifthenelse{\equal{#7}{}}{}{#7} \\
    \textit{#5}
  \end{center}
  \bigskip
}

\renewcommand{\abstract}[1]{%
  \begin{center}
    \begin{minipage}{0.85\textwidth}
      \begin{footnotesize}
        #1
      \end{footnotesize}
    \end{minipage}
  \end{center}
  \bigskip
}

\begin{document}

%
%
%
%
%
%
{  


%

\contribution[The axial charge of  $\Delta $ baryon in QCD]  
{The axial charge of  $\Delta $ baryon in QCD}  
{U.}{Ozdem$^1$}  
{$^1$ Physics Department, Canakkale  Onsekiz Mart University, 17100 Canakkale, Turkey\\
$^2$Physics Department, Middle East Technical University, 06531 Ankara, Turkey}  
{ulasozdem@gmail.com}  
{, A. Kucukarslan$^1$ and  A. Ozpineci$^2$}
%

\abstract{The iso-vector axial-vector form factors of the $\Delta-\Delta$ transition are calculated in the framework of the Light-cone QCD sum rules method. Also, axial charge of $\Delta$ baryon is predicted.
}
%
%
%
\section{Introduction}


The axial charge $g_A$ is an important parameter for low-energy effective theories.
It can also be viewed as an indicator of the phenomenon of spontaneous breaking of chiral symmetry of non-perturbative QCD \cite{Choi:2010ty}.
Form factors describe how hadrons interact with each other and bring forth valuable information about the internal structure of the hadrons.
As the form factors are non perturbative properties, they need to be calculated using
a non-perturbative method. Light cone QCD sum rules~\cite{Braun:1988qv, Balitsky:1989ry, Chernyak:1990ag} is one of the non-perturbative
methods that have been applied to various properties of hadrons including their
form factors.
This method has been rather successful in determining hadron form factors at high $Q^2$.
Using the LCSR, the isovector axial vector form factors of baryons have been calculated \cite{Erkol:2011qh, Kucukarslan:2014bla, Kucukarslan:2014mfa}.
For $\Delta$ baryon isovector axial vector form factors have been studied
using  lattice QCD \cite{Alexandrou:2013opa},
chiral perturbation theory \cite{Jiang:2008we} and
quark models \cite{Choi:2010ty, Theussl:2000sj, Glozman:1997ag, Glantschnig:2004mu}.

\section{The axial form factors }

In the LCSR approach, we consider the following two-point correlation function:
\begin{equation}\label{corrf}
	\Pi_{\mu\nu}(p,q)=i\int d^4 x e^{iqx} \langle 0 |T[\eta_{\mu}^{\Delta}(0) A_\nu^{3}(x)]|\Delta(p,s)\rangle,
\end{equation}
 where $\eta_\Delta(x)$ is an interpolating current for the $\Delta$ baryon which has the form as
  \begin{eqnarray}\label{intf}
		\eta_{\mu}^{\Delta}(0)=&\frac{1}{\sqrt{3}}\epsilon^{abc}[2(u^{aT}(0) C\gamma_\mu d^b(0)) u^c(0)+ (u^{aT}(0)C\gamma_\mu u^b(0))d^c(0)].
	\end{eqnarray}

 The axial transition form factors are defined by the matrix element which can be expressed in terms of four invariant transition factors as \cite{Alexandrou:2013opa};
\begin{eqnarray}
\langle \Delta(p',s')| A_\nu (x)| \Delta(p,s)\rangle = \frac {-i}{2} \overline{\upsilon}^{\alpha}(p',s')\bigg[g_{\alpha\beta}\bigg(g_1^A(q^2) \gamma_\nu \gamma_5 + g_3^A(q^2)\frac{q_\nu \gamma_5}{2M_\Delta}\bigg)  \nonumber \\
+ \frac{q^\alpha q^\beta}{4M_\Delta^{2}}\bigg( h_1^A(q^2)\gamma_\nu \gamma_5+ h_3^A(q^2)\frac{q_\nu \gamma_5}{2M_\Delta}\bigg) \bigg]\upsilon^\beta(p,s)
\end{eqnarray}

where $A^3_\nu(x) = \frac12[( \bar u(x) \gamma_\nu \gamma_5 u(x) -  \bar d(x) \gamma_\nu \gamma_5 d(x)]$ is the isovector-axial vector  current,
$q = p'-p$, $M_\Delta$ is delta mass, $\upsilon_\alpha$ is a Rarita-Schwinger spinor for $\Delta$ baryon.

In order to calculate the correlation function, we need to know matrix element of the three quark operator. This matrix element can be expressed in terms of quark distribution amplitudes (DAs). DAs of the $\Delta$ are
calculated in Refs.~\cite{Carlson:1988gt}. Further details of the sum rules calculations can be found in \cite{Kucukarslan:2014bla}.

Choosing the relevant structures, we determine sum rules for the form factors $g_1^A$, $g_3^A$ as:

\begin{eqnarray*}
	 g_1^A(q^2)  \frac{\lambda_{\Delta}}{M_\Delta-p'^{2}}&=&
	-  \frac{f_\Delta M_\Delta}{\sqrt{3}} \bigg[\int_0 ^{1}dx_2 \frac{1}{(q-px_{2})^2}\int_0^{1-x_2}d_{x_1}
	 4V(x_1,x_2,1-x_1-x_2)\nonumber \\
& &-\int_0 ^{1}dx_3 \frac{1}{(q-px_{3})^2}\int_0^{1-x_3}d_{x_1}[-T+A-2V](x_1,1-x_1-x_3,x_3)\bigg] \nonumber\\
\end{eqnarray*}
\begin{eqnarray}
    g_3^A(q^2)  \frac{\lambda_{\Delta}}{M_\Delta-p'^{2}}&=& -  \frac{f_\Delta M_\Delta}{\sqrt{3}}\bigg[\int_0 ^{1}dx_2 \frac{1}{(q-px_{2})^2}\int_0^{1-x_2}d_{x_1}
                 [2T+4A+8V](x_1,x_2,1-x_1-x_2) \nonumber \\
    & &+\int_0 ^{1}dx_3 \frac{1}{(q-px_{3})^2} \int_0^{1-x_3}d_{x_1} [-3T+3A+2V](x_1,1-x_1-x_3,x_3)\bigg].\nonumber\\
\end{eqnarray}

\section{Results and Conclusions}
In this section, we present our numerical predictions of the axial vector form factors of $\Delta$ baryon.
To obtain the  numerical results, we use the expressions of the $\Delta$ baryon DAs which are studied in \cite{Carlson:1988gt}.
Also, for the value of the residue of the $\Delta$ baryon, $\lambda_\Delta$.
We choose the value as $\lambda_\Delta= 0.038 ~GeV^3$ from analysis of the mass sum rules \cite{Ioffe:1981kw}.

In Fig.1, we plot the dependence of
the form factors $g_1^A{(Q^2)}$ and $g_3^A{(Q^2)}$ on $M^{2}$
for  two fixed values of $Q^2$ and for various values of $s_0$ in the range $2 ~ GeV^2 \leq s_0 \leq  4~GeV^2$.
From these figures, the predictions are quite stable
with respect to variation of the Borel parameter for $s_0=2.5\pm0.5~GeV^2$.

In Fig. 2, we present the form factors $g_1^A$ and $g_3^A$ as a function of $Q^2$.
The qualitative behavior of the form factors agree with our expectations.
Since only the leading twist DAs of the $\Delta$ baryons are known, it is not
enough to determine the other form factors $h_1^A{(Q^2)}$ and $h_3^A{(Q^2)}$.\\

The axial form factor $g_1^{A}(Q^2)$ is the only one that can be extracted directly from the matrix element,
  and we can determine the axial charges at $Q^2 = 0$. However, in our case, the working region
of the LCSR cannot extrapolate to the $Q^2 = 0$ directly. LCSR results more reliable at $Q^2 \geq 1~GeV^2$.
  Therefore, the axial form factor is parameterized in terms of an exponential form
\begin{equation}
	g_{A}(q^2)= g_{A}(0) \exp[-Q^2/M_{A}^2]
\end{equation}
Our results are shown in Table I. For this fit form we have studied three fit regions
$ Q^{2}$ $\geq $ $1~GeV^2 $, $ Q^{2}$ $\geq $ $1.5~GeV^2 $ and $ Q^{2}$ $\geq $ $2~GeV^2 $.

\begin{table}[t]
	\addtolength{\tabcolsep}{2pt}
\begin{tabular}{ccccccc}
				\hline\hline
		 Fit Region~(GeV$^2$) & $g_{A}(0)$ & $M_{A}$~(GeV)& \\[0.5ex]
		\hline
		$ [1.0-10]$ &-3.48& 1.15  &  \\
		$ [1.5-10]$ &-2.64 & 1.24  &  \\
		$ [2.0-10]$ &-2.10 & 1.32  &  \\[1ex]
\hline\hline
	\end{tabular}
\caption{The  values of exponential fit parameters, $g_{A}$ and $M_{A}$ for axial form factors.
The results include the fits from three region. }
	\label{fit_table}
\end{table}

In Table II, we present the different numerical results of the axial charge predicted from other theoretical models. As seen from table,
our result is slightly larger than the result obtained from Lattice QCD \cite{Alexandrou:2013opa}, approximately two times smaller compared to the predictions of ChPT \cite{Jiang:2008we}  and quark models \cite{Theussl:2000sj, Glozman:1997ag, Glantschnig:2004mu}. There is no experimental result yet.
\begin{table}[t]
	\addtolength{\tabcolsep}{2pt}
\begin{tabular}{ccccccc}
				\hline\hline
		   & \cite{Alexandrou:2013opa} &\cite{Jiang:2008we}&\cite{Theussl:2000sj}&\cite{Glozman:1997ag} &\cite{Glantschnig:2004mu}& This Work  \\[0.5ex]
		\hline
		   $g_{A}$ &$- 1.9 \pm 0.1$ & $-4.50$  &  $ - 4.47$& $- 4.48$& $- 4.30$  & $-2.70 \pm 0.6$ \\
		\hline\hline
	\end{tabular}
\caption{Different results from theoretical models which are Lattice QCD \cite{Alexandrou:2013opa}, ChPT \cite{Jiang:2008we}, quark models \cite{Theussl:2000sj, Glozman:1997ag, Glantschnig:2004mu} and also our model. }
	\label{compare}
\end{table}
\section*{Acknowledgments}
The work of U. O. is supported by the Scientific and Technological Research Council of Turkey (TUBITAK).
Many thanks are due to the Organizers for maintaining a stimulating and friendly atmosphere of
this Conference for already more than a twenty years.

\bibliography{refs}

\begin{thebibliography}{10}
\providecommand{\url}[1]{{#1}}
\providecommand{\urlprefix}{URL }
\expandafter\ifx\csname urlstyle\endcsname\relax
  \providecommand{\doi}[1]{DOI \discretionary{}{}{}#1}\else
  \providecommand{\doi}{DOI \discretionary{}{}{}\begingroup
  \urlstyle{rm}\Url}\fi

\bibitem{Choi:2010ty}
K.S. Choi, W.~Plessas, R.~Wagenbrunn, Phys.Rev. \textbf{{\bf D82}}, 014007
  (2010)

\bibitem{Braun:1988qv}
V.M. Braun, I.~Filyanov, Z. Phys. \textbf{{\bf C44}}, 157 (1989)

\bibitem{Balitsky:1989ry}
I.~Balitsky, V.M. Braun, A.~Kolesnichenko, Nucl. Phys. \textbf{{\bf B312}}, 509
  (1989)

\bibitem{Chernyak:1990ag}
V.~Chernyak, I.~Zhitnitsky, Nucl. Phys. \textbf{{\bf B345}}, 137 (1990)

\bibitem{Erkol:2011qh}
G.~Erkol, A.~Ozpineci, Phys.Rev. \textbf{{\bf D83}}, 114022 (2011)

\bibitem{Kucukarslan:2014bla}
A.~Kucukarslan, U.~Ozdem, A.~Ozpineci, Phys. Rev. \textbf{D90}, 054002 (2014)

\bibitem{Kucukarslan:2014mfa}
A.~Kucukarslan, U.~Ozdem, A.~Ozpineci, J. Phys. \textbf{G42}(1), 015001 (2015)

\bibitem{Alexandrou:2013opa}
C.~Alexandrou, E.~Gregory, T.~Korzec, G.~Koutsou, J.~Negele, T.~Sato,
  A.~Tsapalis, Phys.Rev. \textbf{{\bf D87}}, 114513 (2013)

\bibitem{Jiang:2008we}
F.J. Jiang, B.C. Tiburzi, Phys.Rev. \textbf{{\bf D78}}, 017504 (2008)

\bibitem{Theussl:2000sj}
L.~Theussl, R.~Wagenbrunn, B.~Desplanques, W.~Plessas, Eur.Phys.J. \textbf{{\bf
  A12}}, 91 (2001)

\bibitem{Glozman:1997ag}
L.Y. Glozman, W.~Plessas, K.~Varga, R.~Wagenbrunn, Phys.Rev. \textbf{{\bf
  D58}}, 094030 (1998)

\bibitem{Glantschnig:2004mu}
K.~Glantschnig, R.~Kainhofer, W.~Plessas, B.~Sengl, R.~Wagenbrunn, Eur.Phys.J.
  \textbf{{\bf A23}}, 507 (2005)

\bibitem{Carlson:1988gt}
C.E. Carlson, J.~Poor, Phys.Rev. \textbf{{\bf D 38}}, 2758 (1988)

\bibitem{Ioffe:1981kw}
B.~Ioffe, Nucl.Phys. \textbf{{ \bf B188}}, 317 (1981)

\end{thebibliography}
\newpage
\begin{figure}[htp]
\centering
 \subfloat[]{\label{fig:g1Msq.eps}\includegraphics[width=0.5\textwidth]{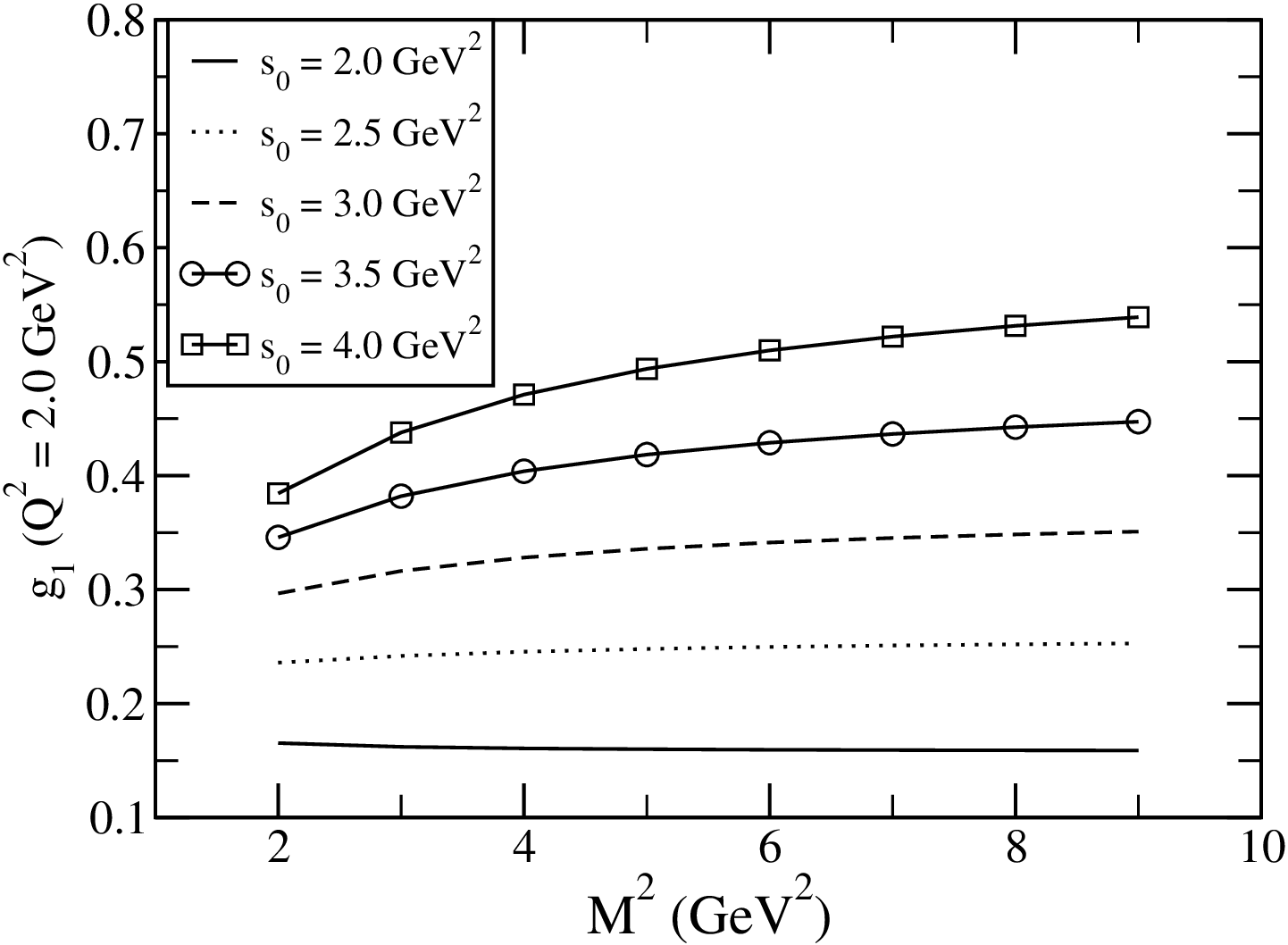}}
  \subfloat[]{\label{fig:g1Msq1.eps}\includegraphics[width=0.5\textwidth]{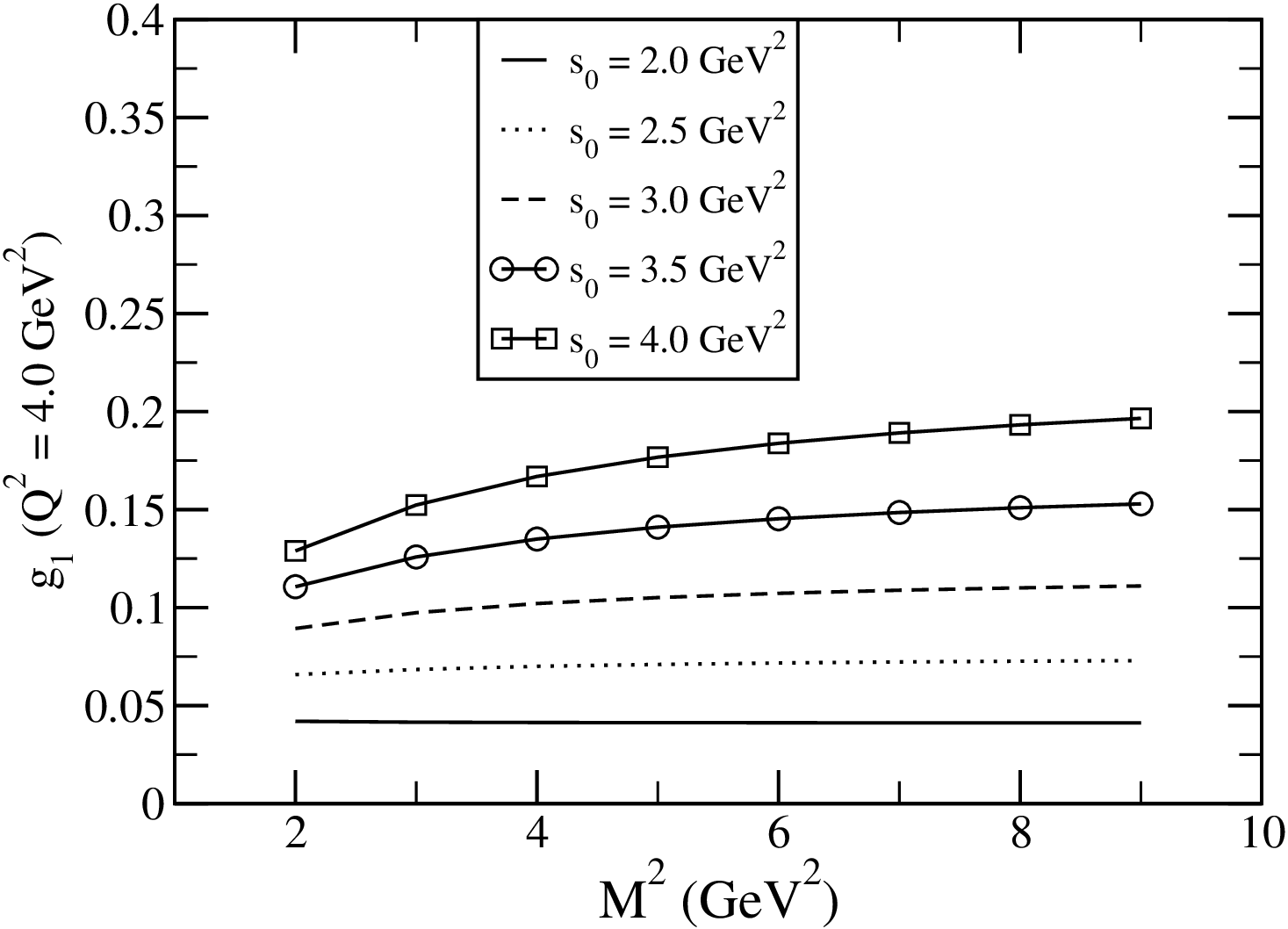}}\\
  \vspace*{0.5cm}
  \subfloat[]{\label{fig:g3Msq.eps}\includegraphics[width=0.5\textwidth]{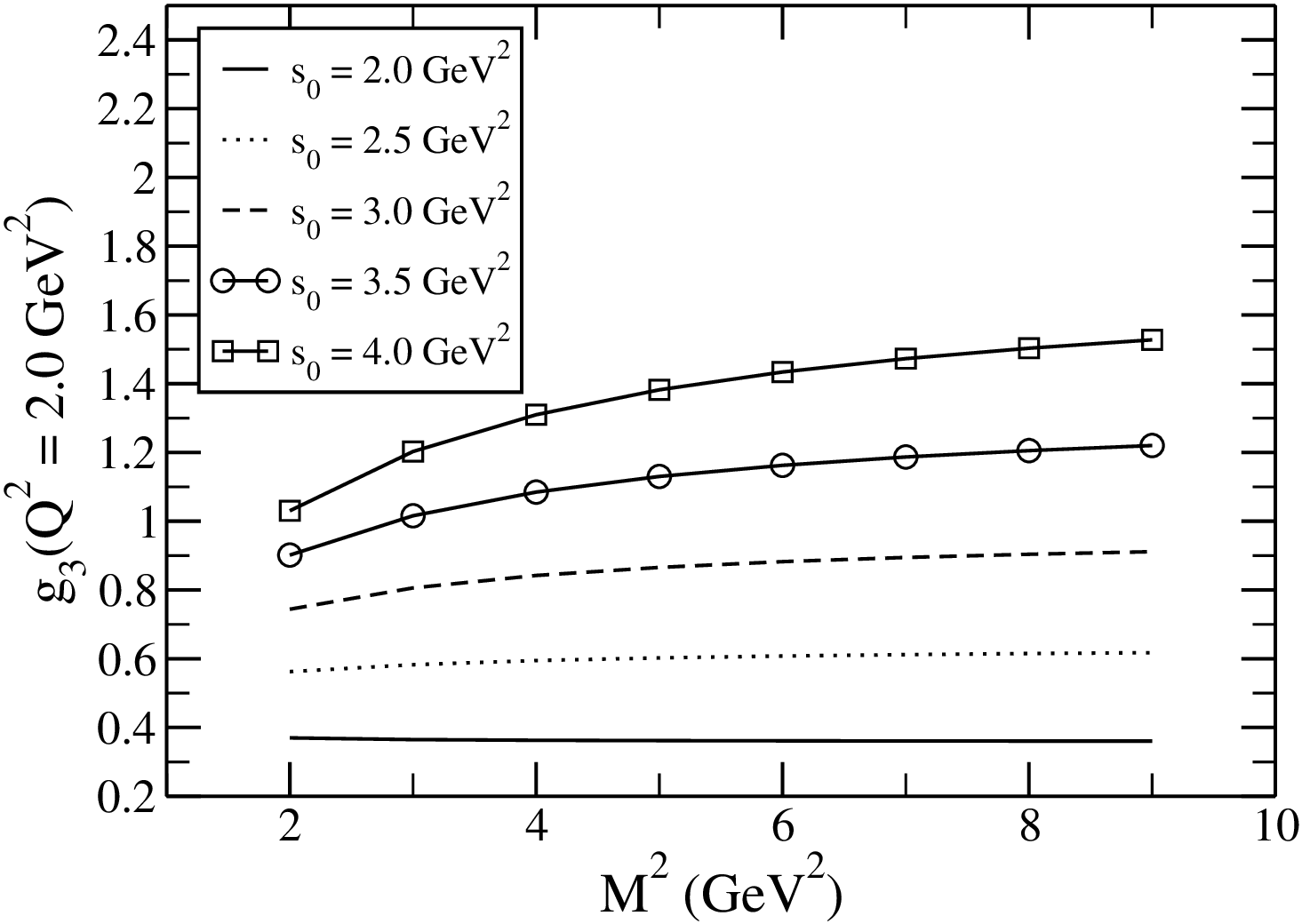}}
  \subfloat[]{\label{fig:g3Msq1.eps}\includegraphics[width=0.5\textwidth]{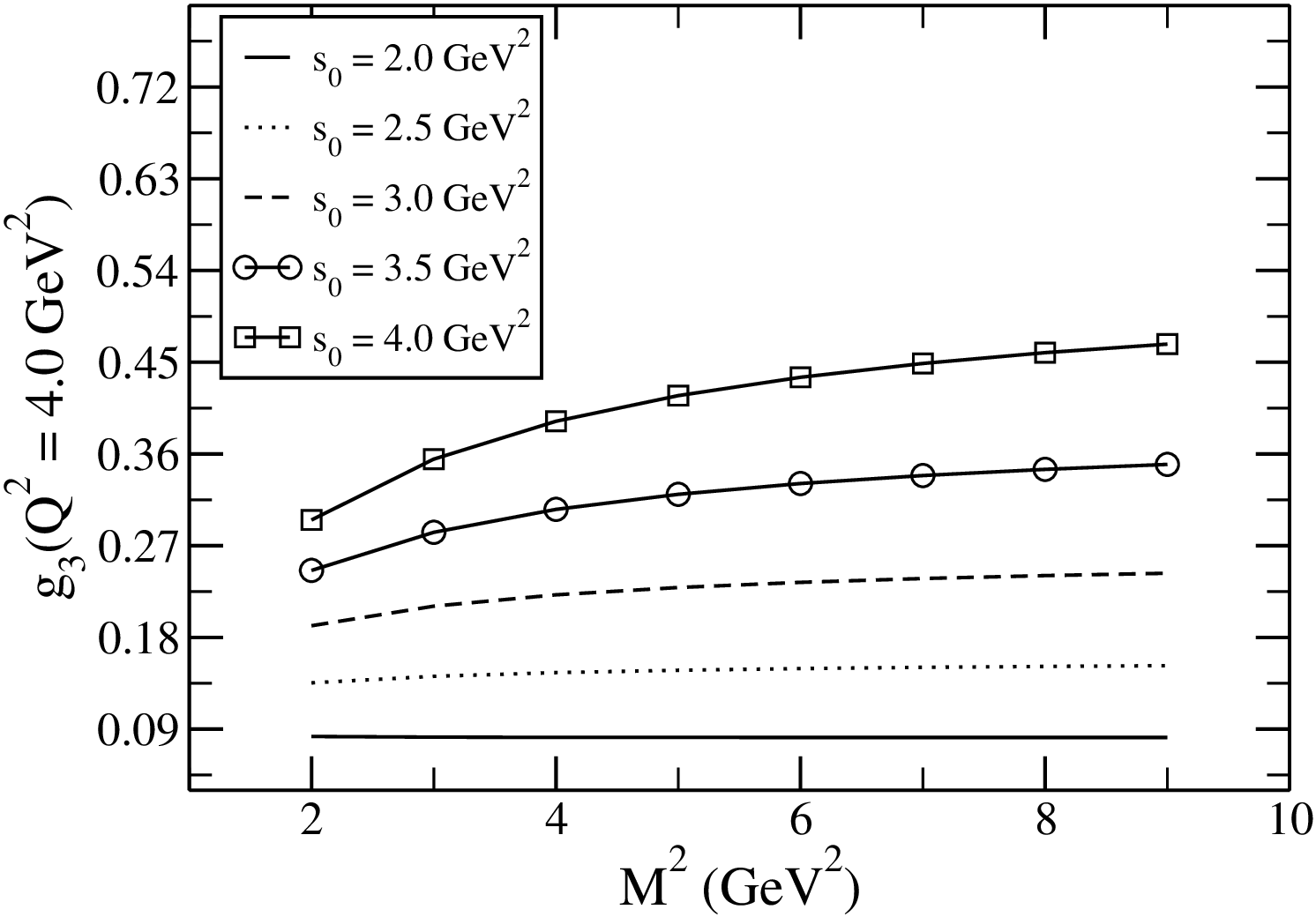}}\\
  \caption{ The dependence of the form factor $g_1^A$ and $g_3^A$ on the Borel parameter squared $M^{2}$
  for the values of the continuum threshold $s_0 = 2.0 ~GeV^2$, $s_0 = 2.5~GeV^2$ $s_0 = 3.0 ~GeV^2$,
  $s_0 = 3.5~GeV^2$ and $s_0 = 4.0~GeV^2$ and at the values $Q^2 = 2.0$ and 4.0 $GeV^2$.
     }
\end{figure}

\begin{figure}[htp]
\centering
 \subfloat[]{\label{fig:g1.eps}\includegraphics[width=0.6\textwidth]{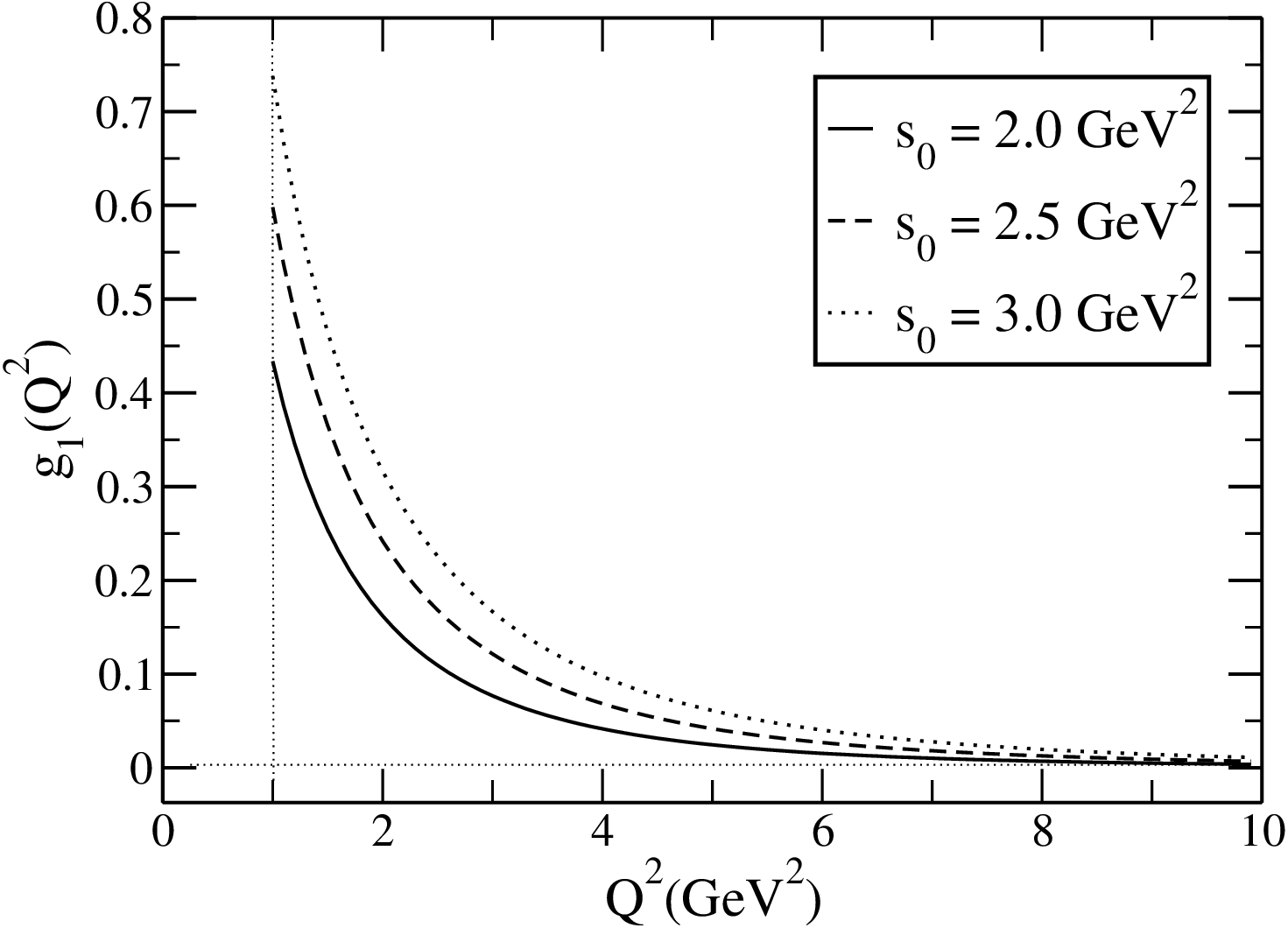}}\\
 \vspace*{0.5cm}
  \subfloat[]{\label{fig:g3.eps}\includegraphics[width=0.6\textwidth]{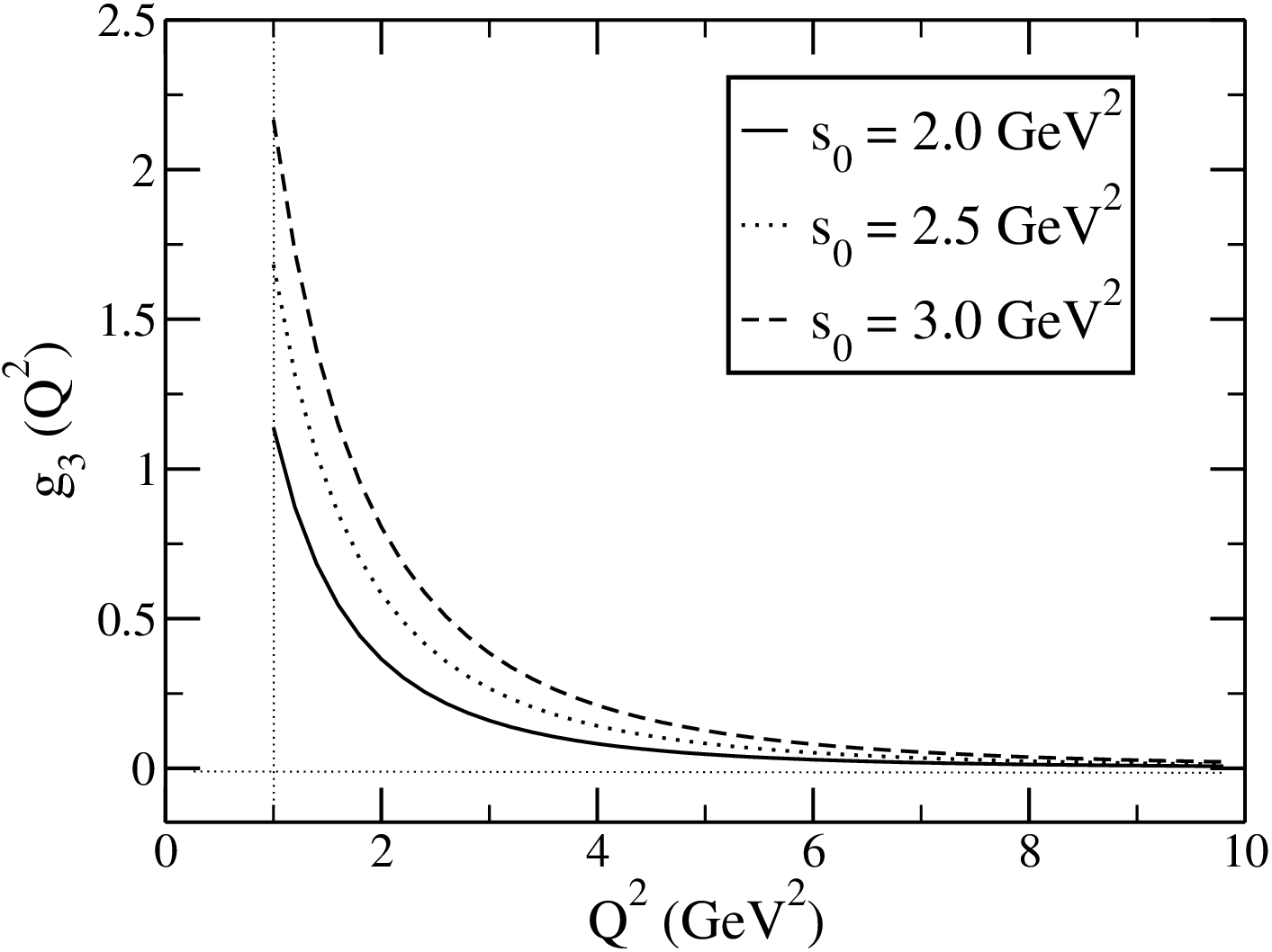}}\\
 \caption{The dependence of the  form factors $g_1^A$ and $g_3^A$ on the $Q^2$  for the values of the continuum threshold
 $s_0 = 2.0 ~GeV^2$, $s_0 = 3.0~GeV^2$, $s_0 = 3.5~GeV^2$ and the Borel parameter $M^{2}=3.0$~$GeV^2$.}
\end{figure}

\end{document}